\documentclass[10pt,twocolumn,letterpaper]{article}

\usepackage{iccv}
\usepackage{times}
\usepackage{epsfig}
\usepackage{graphicx}
\usepackage{amsmath}
\usepackage{amssymb}

\usepackage{caption}
\usepackage{enumitem}
\usepackage{algorithm,algorithmic}

\usepackage[pagebackref=true,breaklinks=true,letterpaper=true,colorlinks,bookmarks=false]{hyperref}

\iccvfinalcopy 

\def\iccvPaperID{29} 

\ificcvfinal\fi

\begin{document}

\title{MGBPv2: Scaling Up Multi-Grid Back-Projection Networks}

\author{Pablo Navarrete~Michelini,\quad Wenbin Chen,\quad Hanwen Liu,\quad Dan Zhu\\
BOE Technology Co., Ltd.\\
{\tt\small \{pnavarre, chenwb, liuhanwen, zhudan\}@boe.com.cn}
}

\ificcvfinal\thispagestyle{empty}\fi

\twocolumn[{%
\renewcommand\twocolumn[1][]{#1}%
\vspace{-3em}
\maketitle
\vspace{-3em}
\begin{center}
    \centering
    \includegraphics[width=\linewidth]{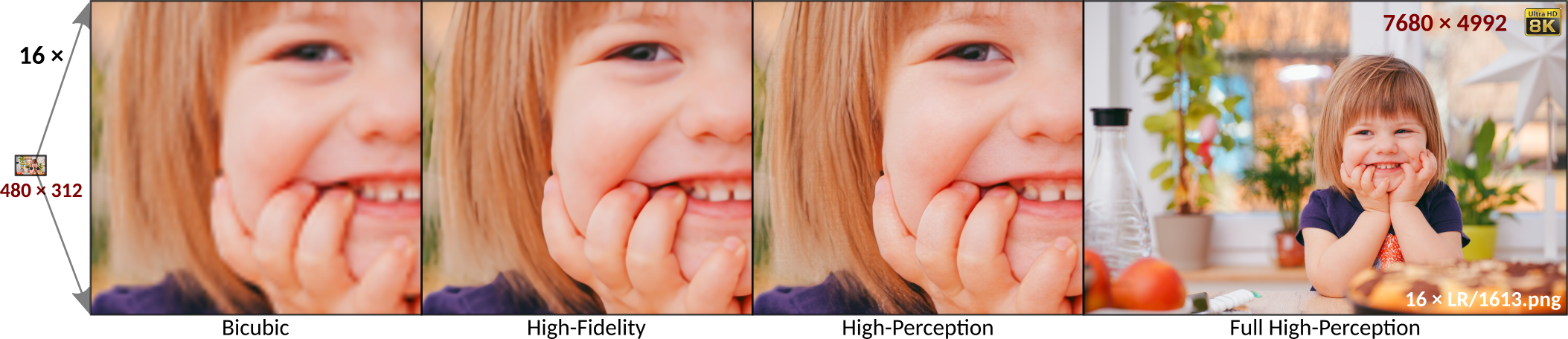}
    \captionof{figure}{
Large upscaling factors are needed to convert standard--definition (SD) to ultra--high--definition (UHD) resolutions. In the task of Extreme Super--Resolution (zooming $16\times$), our proposed MGBPv2 system can achieve a good balance between high fidelity (images close to the original content) and high perceptual quality (images that look real through artificial details).
    }
    \label{fig:teaser}
\end{center}%
}]

\begin{abstract}
   Here, we describe our solution\footnote{Code available at \url{https://github.com/pnavarre/mgbpv2}} for the AIM--2019 Extreme Super--Resolution Challenge, where we won the \textbf{$\boldsymbol{1^{st}}$ place in terms of perceptual quality} (MOS) similar to the ground truth and achieved the $5^{th}$ place in terms of high--fidelity (PSNR). To tackle this challenge, we introduce the second generation of MultiGrid BackProjection networks (MGBPv2) whose major modifications make the system scalable and more general than its predecessor. It combines the scalability of the multigrid algorithm and the performance of iterative backprojections. In its original form, MGBP is limited to a small number of parameters due to a strongly recursive structure. In MGBPv2, we make full use of the multigrid recursion from the beginning of the network; we allow different parameters in every module of the network; we simplify the main modules; and finally, we allow adjustments of the number of network features based on the scale of operation. For inference tasks, we introduce an overlapping patch approach to further allow processing of very large images (e.g. 8K). Our training strategies make use of a multiscale loss, combining distortion and/or perception losses on the output as well as downscaled output images. The final system can balance between high quality and high performance.
\end{abstract}

\section{Introduction}
\begin{figure}
  \centering
  \includegraphics[width=\linewidth]{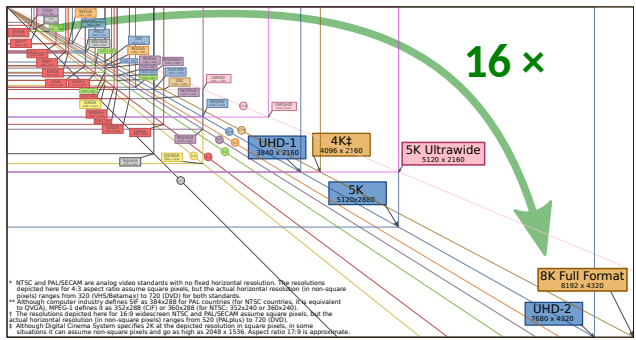}
  \caption{Extreme Super--Resolution is a technology necessary to connect the extremes in the dramatic growth of standard resolutions. Source: \href{https://commons.wikimedia.org/wiki/File:Vector_Video_Standards.svg}{Wikimedia CC BY-SA}.\label{fig:standards}}
\end{figure}

Image upscaling has been studied for decades and remains an active topic of research because of constant technological advances in digital imaging. One scenario where image upscaling is now more demanding arises in digital display technologies, where new standards like BT.2020~\cite{MSugawara_2014a} are introduced. The resolution of digital displays has experienced a tremendous growth over the past few decades, as shown in Figure \ref{fig:standards}. The transition between different formats leads to a challenging problem. On one hand, large amount of digital content still exist in popular old standards such as standard--definition (SD). On the other hand, the latest display technologies (e.g. 4K, 8K and above) are expected to show this content with reasonable quality. Standard upscaling technologies are clearly insufficient for this purpose. While a $2\times$ upscaler maps $1$ input pixel into $4$ output pixels, a $16\times$ upscaler maps $1$ input pixel into $256$ output pixels, which already contain small images. The problem demands advanced solutions that are capable of understanding the image content and filling in these large pieces of images with visually appealing and consistent information. In particular, large upscaling factors are needed to convert SD to ultra high--definition (UHD) resolutions. For example, to upscale from a popular width of $640$ pixels (VGA) to $8,192$ (8K) we need a factor greater than $12\times$. And from $640$ pixels to $10,240$ (10K), which is used by some of the most advanced modern displays~\cite{BOE10K, ran2016p}, we need a $16\times$ factor. Thus, extreme upscaling represents a real problem in current market and it is expected to persist and become even more challenging with the rapid development of new technologies.

The challenge on Extreme Super--Resolution proposed at the 2019 ICCV workshop on Advances in Image Manipulation (AIM--2019) \cite{AIM19_ESR} was the first of its kind to tackle this problem. That is, the task of increasing the resolution of an input image with a magnification factor $16\times$ based on a set of prior examples of low and corresponding high resolution images. The challenge provided a novel DIVerse up to 8K resolution images dataset (DIV8K) with a large diversity of content. Two targets were proposed:
\begin{itemize}[leftmargin=*]
    \item \textbf{Track 1: Fidelity}, the aim is to obtain a solution capable to produce high resolution results with the best fidelity (PSNR) to the ground truth.
    \item \textbf{Track 2: Perceptual}, the aim is to obtain a solution capable to produce high resolution results with the best perceptual quality similar to the ground truth.
\end{itemize}
These two targets are motivated by the so--called \emph{perception--distortion} trade--off established by Blau and Michaeli in \cite{Blau_2018_CVPR}. That is, both targets cannot be achieved at the same time, one must compromise perceptual quality to improve fidelity and vice versa. The trade--off motivated the Perceptual Image Restoration and Manipulation (PIRM) workshop at ECCV--2018~\cite{PIRM-SR} that strongly confirmed and gave further insight into this fundamental principle.

In the Extreme--SR competitions, we identified two major challenges: performance scalability, due to the large size of the output images; and content scalability, due to the diverse scale of image content. Our solution was built upon the MultiGrid BackProjection (MGBP) network proposed in \cite{PNavarrete_2019a, G-MGBP}, combining both the iterative backprojection and multigrid algorithms. First, the \textbf{iterative backprojection algorithm provides performance}. With solid ground in the classical theory of image super--resolution (SR), its deep--learning versions have succeeded in competitions. Examples include DBPN~\cite{DBPN2018} (which won the NTIRE--2018 challenge on $8\times$ SR and the Region--2 track of PIRM--SR 2018\cite{PIRM-SR}) and MGBP~\cite{G-MGBP} (which won the $2^{nd}$ place in the Region--3 track, i.e. best perception, of PIRM--SR 2018\cite{PIRM-SR}). Second, the \textbf{multigrid algorithm provides scalability}. Here, it is worth noting that \emph{multigrid} is one of the most popular algorithms in the field of PDE and numerical solvers\cite{UTrottenberg_2000a}, but its name is often misinterpreted in other fields as merely meaning ``multilevel''. Multigrid is one of the few known methods that can solve numerical systems of linear equations with optimal computational complexity (linear in the number of unknowns). The MGBP algorithm does not work as a linear solver, but borrows the multigrid recursion to scale the iterative backprojection algorithm. The recursion unfolds into a convenient structure, in which the computational complexity grows linearly with the image size. This is visible in the diagram of the system (see for example Figure \ref{fig:mgbpv2}), where the number of modules is maximum at the lowest resolution and decreases exponentially as the resolution increases.

Despite its convenient properties, the original formulation of MGBP does not scale well for extreme upscaling factors. This is due to two problems: first, the small number of parameters in MGBP limits its performance on image quality compared to large models; second, the recursive structure forces the number of network features to remain constant along scales. Our main contribution is to redesign MGBP to fix these problems. We summarize our contributions as follows:
\begin{itemize}[leftmargin=*]
    \item We propose to make full use of the multigrid recursion from the beginning of the network.
    \item We propose to simplify the main modules and allow different parameters for every instance in the network.
    \item We propose a strategy to merge overlapping patches in inference to further allow processing of very large images.
    \item We propose training strategies at multiple scales, combining distortion and/or perception losses on the output as well as downscaled output images.
\end{itemize}

\section{System Architecture}

\begin{figure*}
  \centering
  \includegraphics[width=\linewidth]{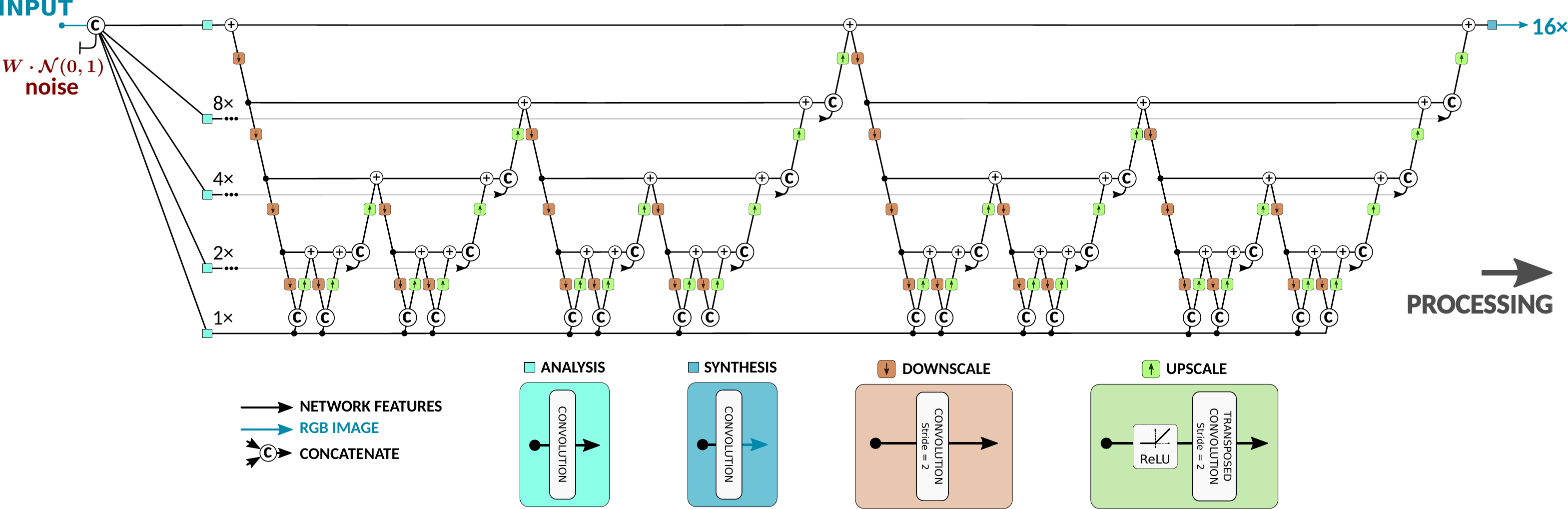}
  \caption{Diagram of the MultiGrid BackProjection version 2 (MGBPv2) network unfolded from Algorithm \ref{alg:mgbp2} with $\mu=2$. \label{fig:mgbpv2}}
\end{figure*}
\begin{algorithm*}
    \caption{Multi--Grid Back--Projection version 2 (MGBPv2)} \label{alg:mgbp2}
    \begin{tabular}{ll}
        \textbf{MGBPv2}$(X,W,\mu,L)$: & $\boldsymbol{BP^{\mu}_{k}}(u\;|\;y_1,\ldots,y_{k-1}; \text{tag}_1,\ldots,\text{tag}_{k-1})$: \\

        \resizebox{.45\textwidth}{!}{
        \begin{minipage}{.5\textwidth}
            \begin{algorithmic}[1]
                \REQUIRE Input image $X$.
                \REQUIRE Steps $\mu$, levels $L$ and noise amplitude $W$.
                \ENSURE Output image $Y$.

                \STATE $x = [X, \; W\cdot\mathcal{N}(0,1)]$
                \FOR{$k = 1,\ldots,L$}
                    \STATE $y_k = \text{Analysis}_{k}(x)$
                \ENDFOR
                \FOR{$k = 1,\ldots,L$}
                    \STATE $\text{tag}_k = 0$
                \ENDFOR
                \STATE $y = BP^{\mu}_L(y_L\;|\;y_1,\ldots,y_{L-1}; \text{tag}_1,\ldots,\text{tag}_{L-1})$
                \STATE $Y = \text{Synthesis}(y)$
            \end{algorithmic}
        \end{minipage}
        }

        &

        \resizebox{.45\textwidth}{!}{
        \begin{minipage}{0.53\textwidth}
            \begin{algorithmic}[1]
                \REQUIRE Input image $u$, level index $k$, steps $\mu$.
                \REQUIRE Images $y_1,\ldots,y_{k-1}$ and $\text{tag}_1,\ldots,\text{tag}_{k-1}$ (for $k>1$).
                \ENSURE Image $out$

                \STATE $out = u$
                \IF{$k > 1$}
                    \FOR{$s = 1,\ldots,\mu$}
                        \STATE $\text{tag}_{k-1} = s$
                        \STATE $LR = \text{Downscale}_{tag}(out)$
                        \STATE $c = BP^{\mu}_{k-1}(LR\;|\;y_1,\ldots,y_{k-2}; \text{tag}_1,\ldots,\text{tag}_{k-2})$
                        \STATE $out = out + \text{Upscale}_{tag}([\;y_{k-1}, c\;])$
                    \ENDFOR
                \ENDIF
            \end{algorithmic}
            \end{minipage}
        }
    \end{tabular}
\end{algorithm*}

\begin{figure*}
  \centering
  \includegraphics[width=.8\linewidth]{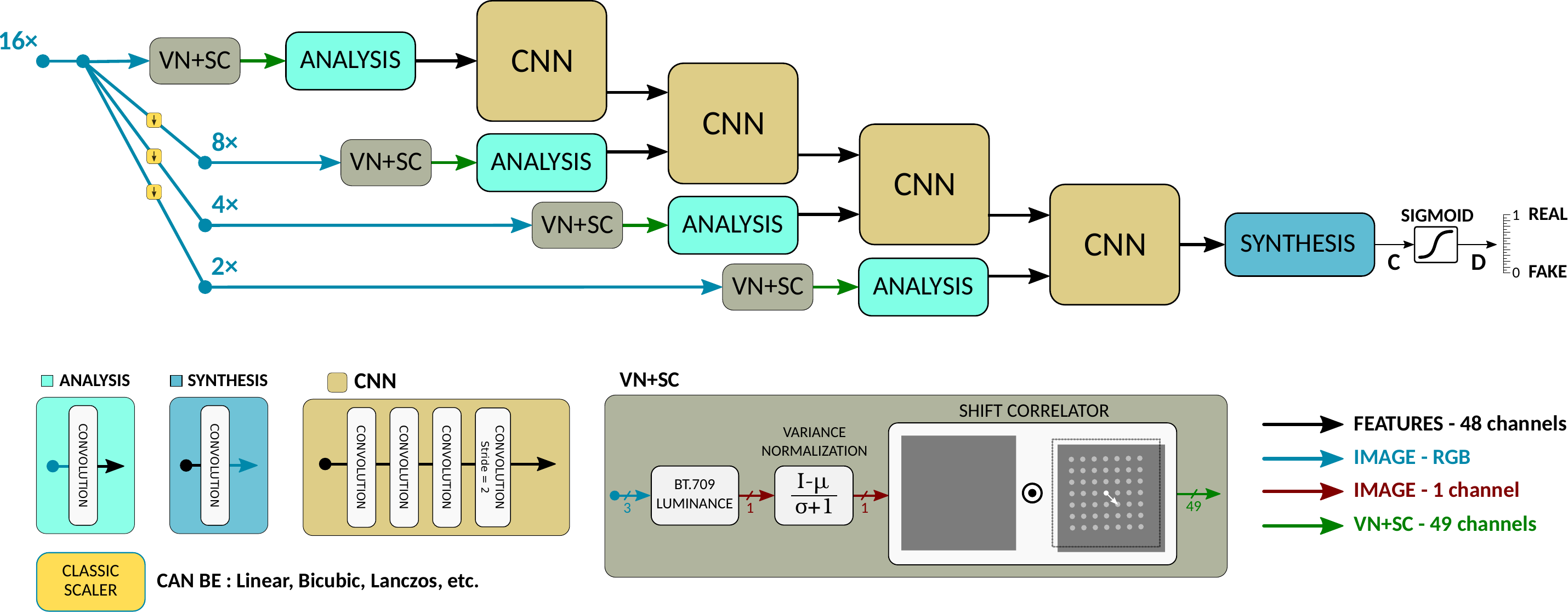}
  \caption{Discriminator system used for adversarial training. The high resolution input image is downsized with standard bicubic downscalers to enter the system at different scales. CNN modules do not share parameters. \label{fig:discr}}
\end{figure*}

 Without loss of generality, we tackle the image enhancement problem with an input resolution equal to the output resolution. For the super--resolution task we input the low resolution image upscaled $16\times$ using a bicubic method. This helps to make the system become more general for applications and simplifies the process of generating pairs of input/output patches during training.

We propose the MGBPv2 algorithm shown in Algorithm \ref{alg:mgbp2}. In Figure \ref{fig:mgbpv2} we display the diagram of the algorithm unfolded for $\mu=2$ and $L=5$. The \emph{Analysis} and \emph{Synthesis} modules covert an image into features space and vice--versa using single convolutional layers. The \emph{Upscaler} and \emph{Downscaler} modules are composed of single strided (transposed and conventional) convolutional layers. An important observation in Algorithm \ref{alg:mgbp2} that makes a significant difference with the original MGBP is the use of a \texttt{tag} label to differentiate each \emph{Upscaler} and \emph{Downscaler} module. This simple trick makes every \emph{Upscaler} and \emph{Downscaler} module different in terms of parameters and hyper--parameters. In particular, we now set the number of features in convolutional layers to $256$, $192$, $128$, $92$, $48$ and $9$ from low to high resolution levels, allowing large images to be processed at high resolutions. The system can be initialized by a \emph{dry run} of Algorithm \ref{alg:mgbp2} where no computation is performed and modules are defined with their correspondent \texttt{tag} labels.

For the Perceptual track we used a single MGBPv2 system configured with filters of $3\times 3$. For the Fidelity track we used an ensemble solution that sums the outputs of $3$ systems with identical configuration except for different filter sizes: $3\times 3$, $5\times 5$ and $7\times 7$.

We concatenate a single noise channel, $W\cdot\mathcal{N}(0,1)$, to the bicubic upscaled input. The amplitude of the noise, $W\in\mathbb{R}$, is set to zero in the Fidelity track and its purpose is to help in the adversarial training for the Perceptual track. Following the design of the generative MGBP algorithm in \cite{G-MGBP}, the noise activates and deactivates the generation of artificial details. As opposed to \cite{G-MGBP} we now generate a single noise channel at the highest resolution, that moves along with the input image to enter the network at different scales by using \emph{Analysis} modules with different strides. In \cite{G-MGBP}, different i.i.d. noise channels are generated at each resolution. This change becomes necessary during inference, where we merge the output of several overlapping patches and therefore we need to use noise images that overlap in the same way as the input image.

For the discriminator in adversarial training we use the system shown in Figure \ref{fig:discr}. Here, we use $4$--layer sequential CNNs with $3\times 3$ filters and stride $1$ except for the last layer that uses stride $2$ to downscale the features. The system resembles the design of the discriminator in \cite{G-MGBP} with two major differences: first, the system in Figure \ref{fig:discr} is not recursive (all CNNs have different parameters); and second, the system uses a single high--resolution input that is downsized using fixed bicubic downscalers to enter at different levels of the system. We use the same \emph{variance normalization and shift correlator} (VN+SC) module from \cite{G-MGBP}.

\section{Training Strategies}

We denote $Y_{W=0}$ and $Y_{W=1}$ the outputs of the generator architecture using noise amplitudes $W=0$ and $W=1$, respectively. Let $S_f$ represent a bicubic downscaler that reduces the resolution by a factor $f$. We know that the original image was downscaled with $S_{16}$. Then, our training strategies in terms of loss functions are defined follows:

\textbf{Fidelity Track:} We use $\mathcal{L}^{L1}(x,y)=\mathbb{E}\left[|x-y|\right]$ at several resolutions to define the total loss function:
\begin{align}
    \mathcal{L}(Y, X; \theta) = \; & \mathcal{L}^{L1}(Y_{W=0}, X) \; + \nonumber \\
                                   & \mathcal{L}^{L1}(S_2(Y_{W=0}), S_2(X)) \; + \nonumber \\
                                   & \mathcal{L}^{L1}(S_4(Y_{W=0}), S_4(X)) \; + \nonumber \\
                                   & \mathcal{L}^{L1}(S_8(Y_{W=0}), S_8(X)) \; + \nonumber \\
                                   & \mathcal{L}^{L1}(S_{16}(Y_{W=0}), S_{16}(X)) \;.
\end{align}
After every epoch we evaluate the current model using $\mathcal{L}^{L2}(x,y)=\mathbb{E}\left[(x-y)^2\right]$ in the validation metric:
\begin{align}
    \mathcal{V}(Y; \theta) = \mathcal{L}^{L2}(Y_{W=0}, X) \;.
\end{align}
We recorded the best models according to this metric (directly related to PSNR) during the training process and we performed further manual tests on larger images afterwards.

\begin{figure*}
  \centering
  \includegraphics[width=\linewidth]{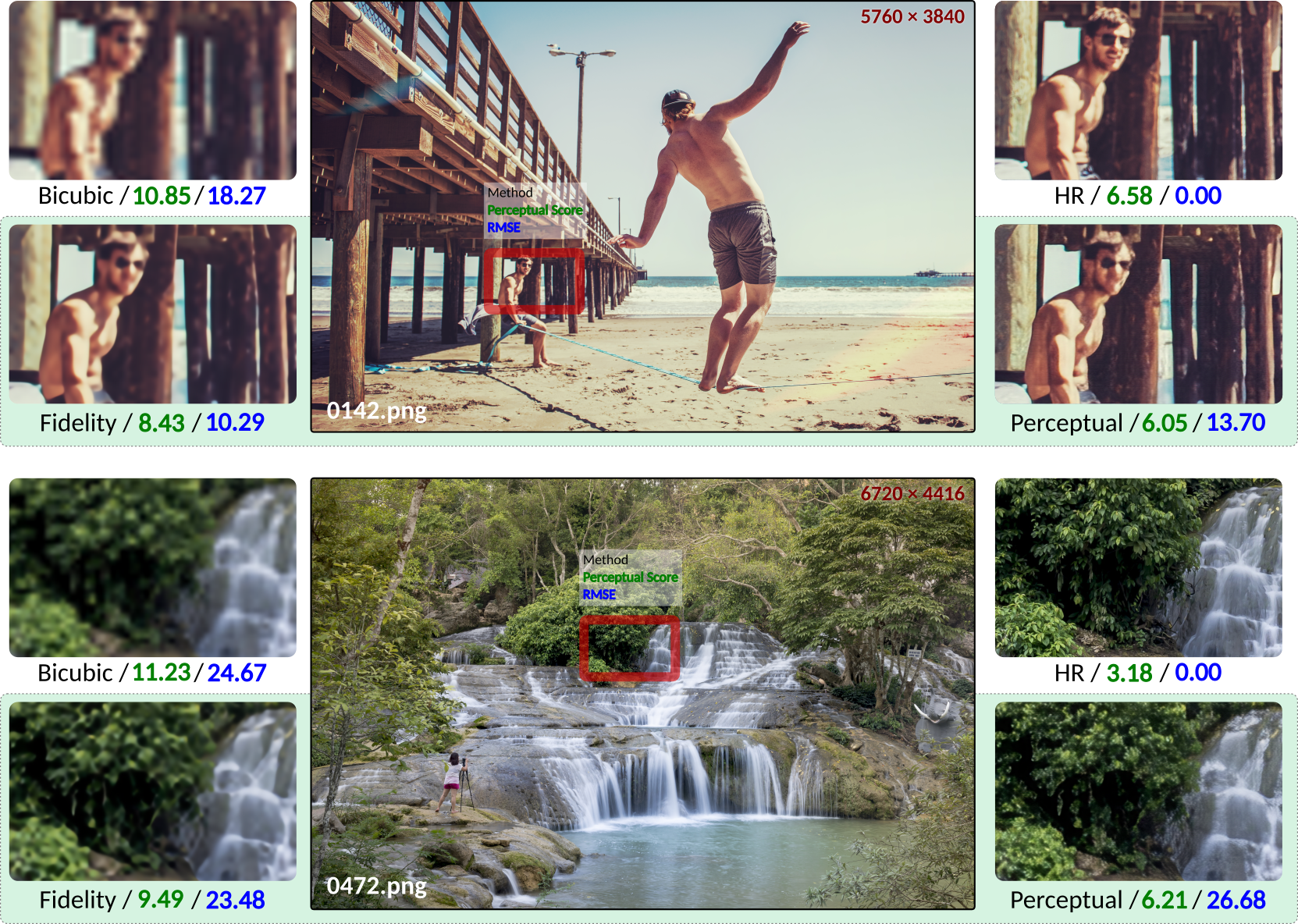}
  \caption{Example outputs and performance metrics for images in the DIV8K training set that were used only for validation. \label{fig:comparison1}}
\end{figure*}

\textbf{Perceptual Track:} We follow the design of multi--scale loss from the generative MGBP in \cite{G-MGBP} with some simplifications to improve training performance. One significant change is in the adversarial loss. We changed the standard GAN used in \cite{G-MGBP} to a Relativistic GAN~\cite{jolicoeur2018relativistic} following the recommendation in \cite{wang2018esrgan}. Our total loss is given by:
\begin{align}
    \mathcal{L}(Y, X; \theta) = \; & 0.001 \cdot \mathcal{L}^{RSGAN}_G(Y_{W=1}) \; + \nonumber \\
                                & 10 \cdot \mathcal{L}^{L1}(S_{16}(Y_{W=1}), S_{16}(X)) \; + \nonumber \\
                                & 0.1 \cdot \mathcal{L}^{CX}(Y_{W=1}, X) \; + \nonumber \\
                                & 10 \cdot \mathcal{L}^{L1}(Y_{W=0}, X) \; + \nonumber \\
                                & 10 \cdot \mathcal{L}^{L1}(S_{16}(Y_{W=0}), S_{16}(X)) \;.
\end{align}
Here, $\mathcal{L}^{CX}$ is the \emph{contextual loss} as defined in \cite{mechrez2018contextual} using features from \emph{conv3--4} of a VGG--19 network as suggested in \cite{mechrez2018Learning}. Ablation tests in \cite{G-MGBP} have shown the effectiveness of this loss function to improve perceptual quality while maintaining a reasonable level of distortion. Next, the Relativistic GAN loss follows the definition in \cite{jolicoeur2018relativistic}, given by:
\begin{align}
\mathcal{L}_D^{RSGAN} = & -\mathbb{E}_{(R,F)}\left[ \log (\text{sigmoid}(C(R)-C(F))) \right] \;,\nonumber \\
\mathcal{L}_G^{RSGAN} = & -\mathbb{E}_{(R,F)}\left[ \log (\text{sigmoid}(C(F)-C(R))) \right] \;.
\end{align}
Here, $C$ is the output of the discriminator before the sigmoid function, as shown in Figure \ref{fig:discr}. And $R$ and $F$ are the sets of real and fake inputs to the discriminator, given by:
\begin{align}
    F = & \left\{ Y_{W=1}, S_{2}(Y_{W=1}), S_{4}(Y_{W=1}), S_{8}(Y_{W=1}) \right\} \;,  \nonumber \\
    R = & \left\{ X, S_{2}(X), S_{4}(X), S_{8}(X) \right\} \;.
\end{align}
After every epoch we evaluated the current model according to the validation metric based on the NIQE\cite{NIQE_2013} index:
\begin{align}
    \mathcal{V}(Y; \theta) = \mathbb{E}\Big[ & NIQE(Y_{W=1}) + NIQE(S_2(Y_{W=1})) \; + \nonumber \\
                                             & NIQE(S_4(Y_{W=1})) \Big] \;.
\end{align}
This metric works as a simple rule to help identify models that generate realistic images in the full resolution, as well as two levels of resolutions below. In other words, we want output images to look real as seen in a display device both from close and far away distances. We performed further subjective tests on larger images to select the best models.

\section{Inference Strategies}

\begin{figure}
  \centering
  \includegraphics[width=.8\linewidth]{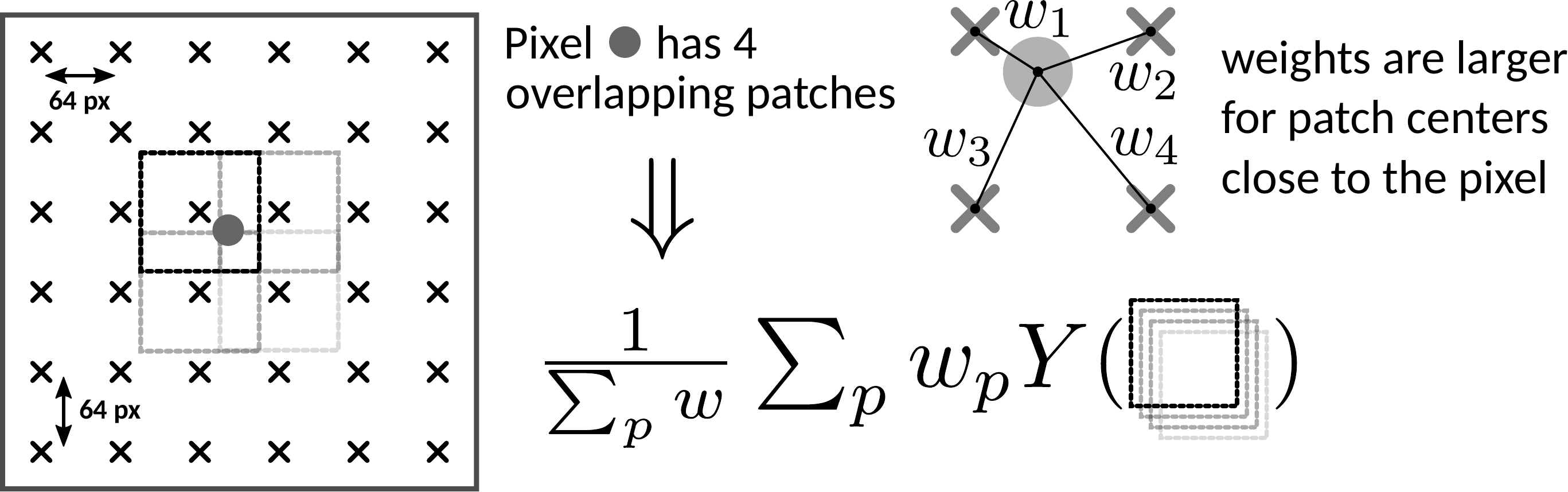}
  \caption{We average overlapping output patches to allow processing arbitrarily large images during inference.\label{fig:overlapping}}
\end{figure}

To upscale large images we propose a patch based approach in which we average the output of overlapping patches taken from the bicubic upscaled input. First, we divide input images into overlapping patches (of same size as training patches) as shown in Figure \ref{fig:overlapping}; second, we multiply each output by a weight decreasing as the distance to the center of a patch increases (we recommend a Hamming window \cite{harris1978use}); and third, we average the results. In our experiments we used overlapping patches separated by $64$ and $128$ pixels in vertical and horizontal directions for the Perceptual and Fidelity tracks, respectively. The weighted average helps to avoid blocking artifacts.

\begin{figure}
  \centering
  \includegraphics[width=.8\linewidth]{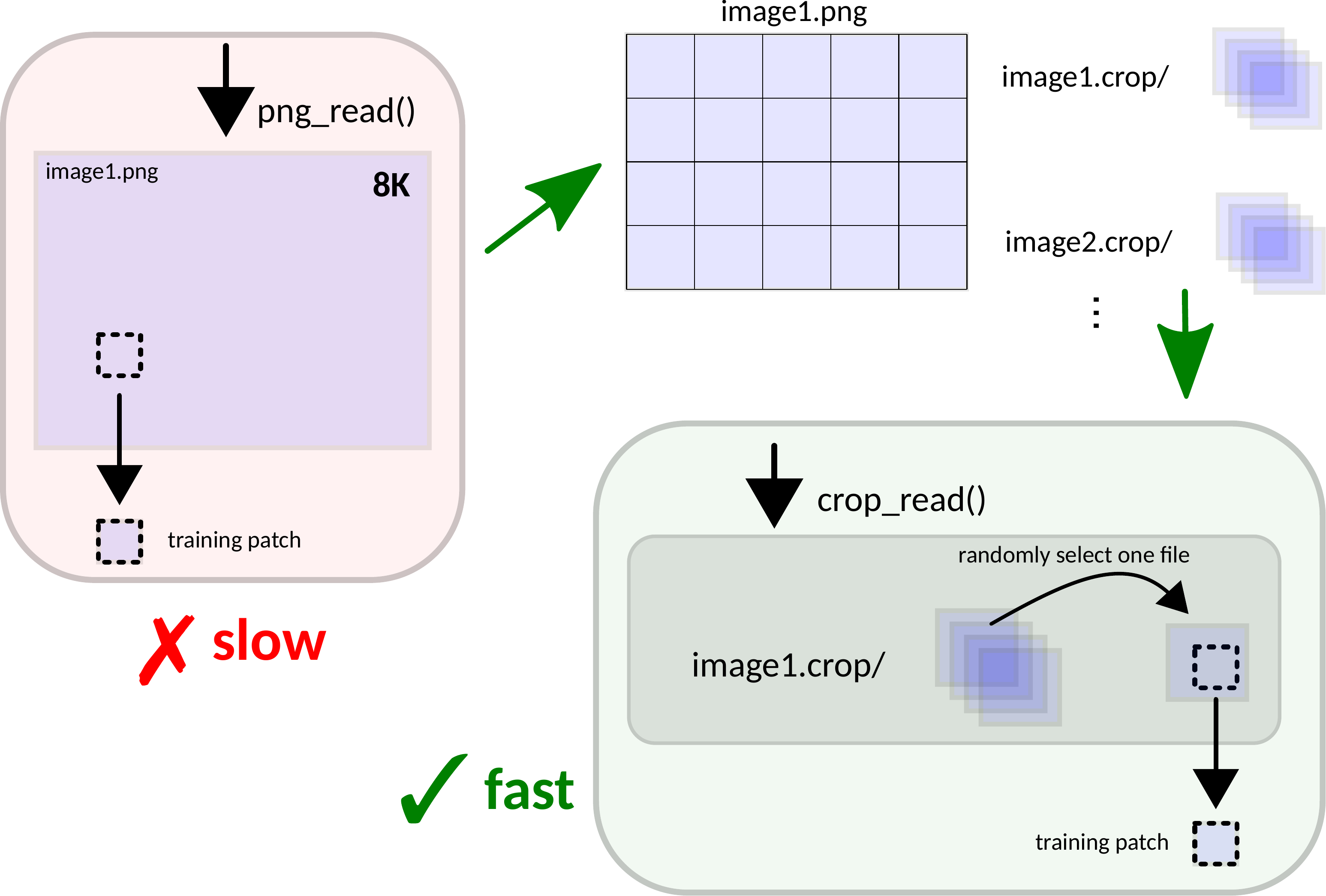}
  \caption{To improve the reading speed in the training process, we replaced images by folders with pre--selected image patches of similar but larger size than training patches.\label{fig:crop}}
\end{figure}

Another problem arise in the training process because reading several 8K images to form a minibatch becomes much slower than running the training step on the minibatch. A popular solution is to save image arrays in a binary format that is much faster to decode (e.g. EDSR~\cite{Lim_2017_CVPR_Workshops}). This solution is not practical here since the amount of space required for the whole dataset exceeds reasonable storage resources. We propose a solution by pre--processing the dataset, creating folders for every image in the training set and saving overlapping patches as image files inside each folder, as shown in Figure \ref{fig:crop}. During training we: first, select images randomly; second, enter the directory and randomly select an image file; third, open the image file and randomly select a patch inside (having similar sizes).

\section{Experiments}
\textbf{Settings:}
We run training tasks on Tesla V100 GPUs and testing on Titan X GPUs.  We used $100$ images from the training set for validation during training. We trained using patch size $667\times 667$ (Fidelity Track) and $767\times 767$ (Perceptual Track) at high resolution. We trained our systems using a Quasi-hyperbolic Adam optimizer~\cite{ma2018quasi} with constant learning rate $10^{-4}$, and changed to Lookahead~\cite{zhang2019lookahead} combined with RAdam~\cite{liu2019radam} in later stages of development. We set the minibatch size to $16$ patches. After every epoch we run the validation metrics described above on the center patch of our $100$ validation images.

\textbf{Performance:} We report a running time of $47.11$ [s] to process a standard 8K image ($7689\times 4320$) on a Titan X (Maxwell) GPU, without using the overlapping patch approach. It takes $1.42 [s]$ in average to process $1000\times 1000$ images used in the validation stage. For our submissions we use the system in the slowest mode, which uses overlapping patches with small distances between the center of the patches ($64$ or $128$ pixels). By using this approach it takes $4.16 [s]$ to process a $1000\times 1000$ image using $667\times 667$ overlapping patches with centers separated by $128$ pixels (used in the Fidelity track). It takes $2.06 [s]$ to process a $1000\times 1000$ image using $767\times 767$ overlapping patches with centers separated by $64$ pixels (used in the Perceptual track). Overall, in both the Fidelity and Perceptual tracks, it takes approximately $16$ minutes to generate one 8K output image using the slowest mode. The overlapping patch approach gives marginal improvements in PSNR and also allows the network to run on devices with low memory resources. The settings used for the competition are purposely not practical for applications as they focus exclusively on image quality. Nevertheless, our empirical tests show that by using larger distances between overlapping patches and multiple GPU devices we can output an 8K image in less than $10$ seconds without significant loss in quality.
\begin{figure}
  \centering
  \includegraphics[width=\linewidth]{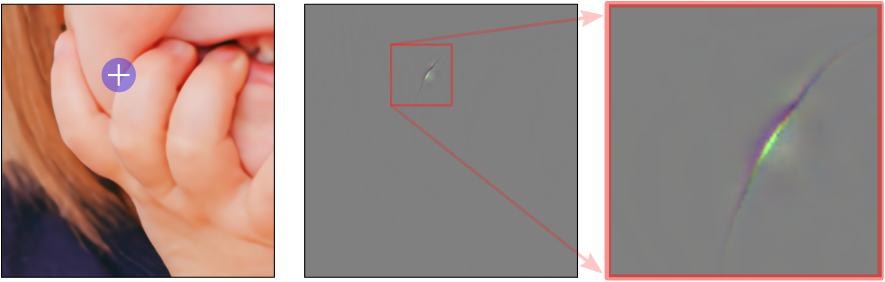}
  \includegraphics[width=\linewidth]{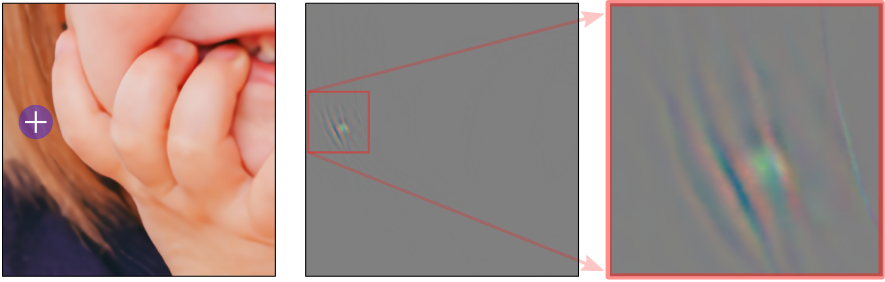}
  \includegraphics[width=\linewidth]{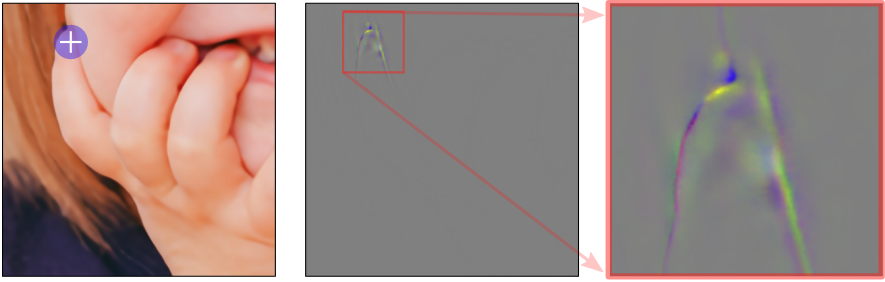}
  \includegraphics[width=\linewidth]{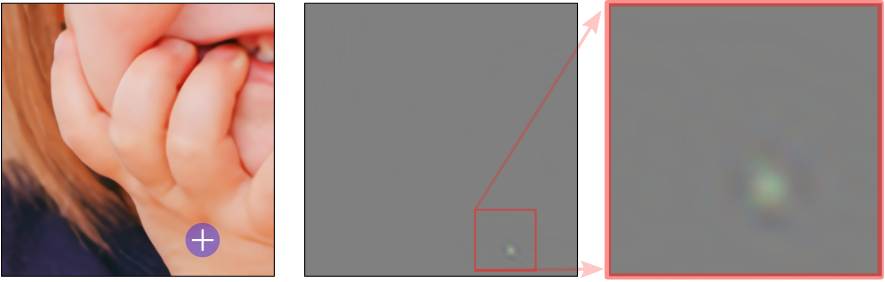}
  \caption{Deep filter visualization (DFV)\cite{PNavarrete_2019a, LinearScopes} experiments on our model for the Fidelity track using a patch of size $767\times 767$. The model shows good knowledge of the geometry and large receptive fields. \label{fig:visualization}}
\end{figure}
\begin{figure*}
  \centering
  \includegraphics[width=\linewidth]{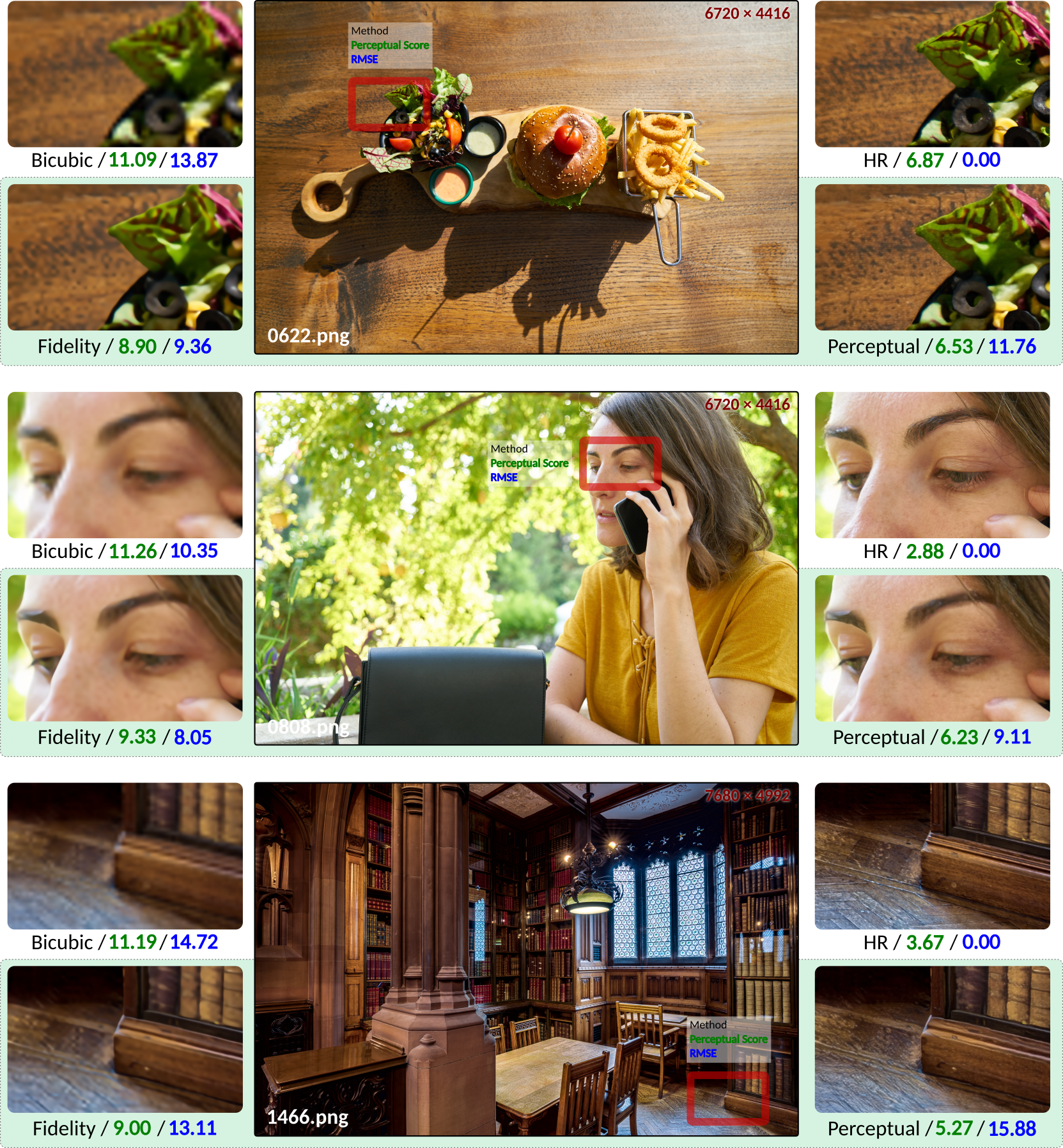}
  \caption{Example outputs and performance metrics for images in the DIV8K training set that were used only for validation. \label{fig:comparison2}}
\end{figure*}
\begin{figure*}
  \centering
  \includegraphics[width=\linewidth]{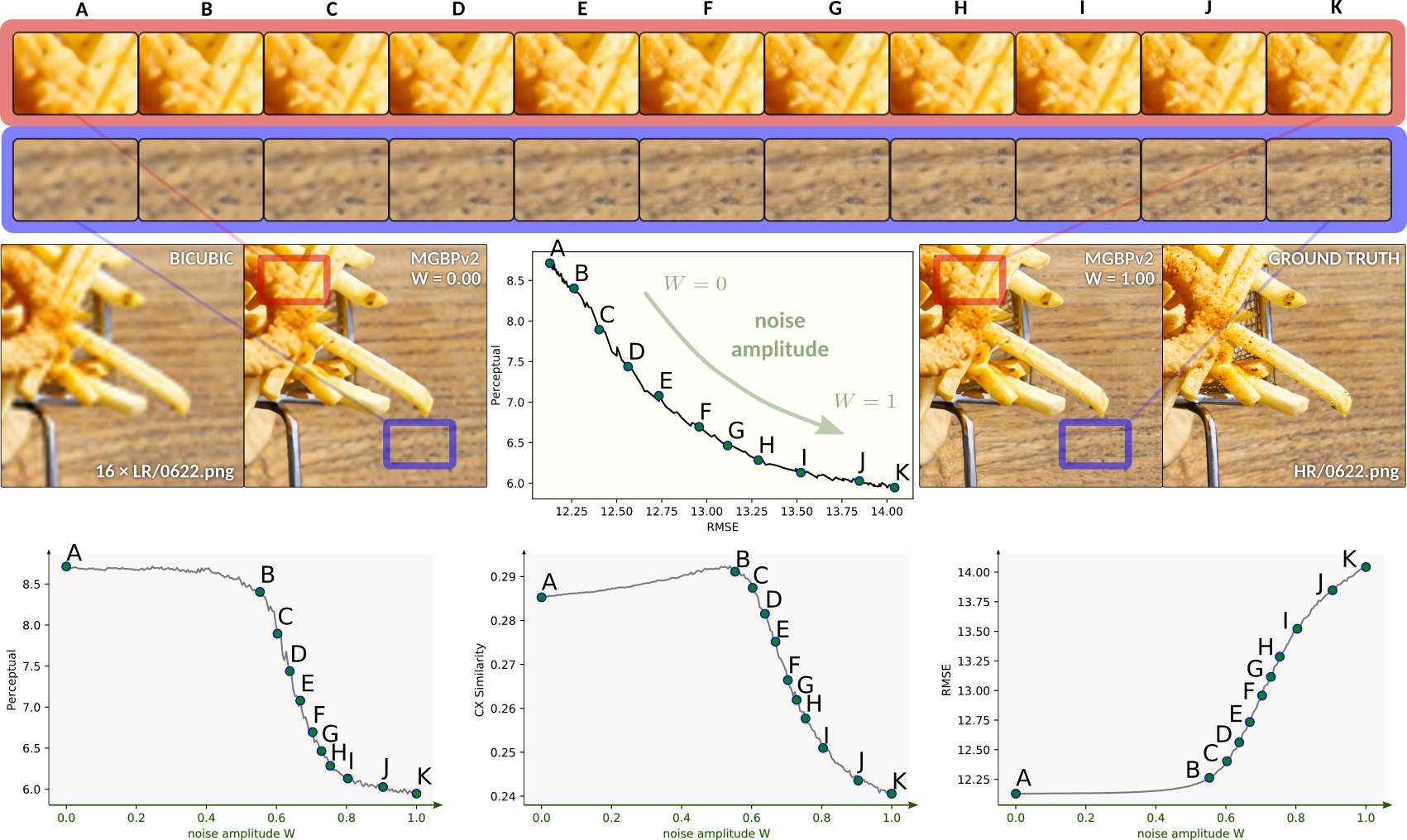}
  \caption{Evolution of perceptual and fidelity metrics when moving the input noise amplitude from $W=0$ to $W=1$. \label{fig:interpolation}}
\end{figure*}

\textbf{Challenge Results:}
The Fidelity track of the Extreme-SR AIM 2019 challenge had $70$ participants, with $9$ finalists submitting results for the test stage. Our results obtained the \textbf{$\boldsymbol{5^{th}}$ place}, with an average PSNR of $26.59$ dB in the full output images of the test set. This is, $0.2$ dB below the top score winning this track.

The Perceptual track of the Extreme--SR AIM 2019 challenge had $52$ participants, with $6$ finalists submitting results for the test stage. Our results obtained the \textbf{$\boldsymbol{1^{st}}$ place} in terms of a subjective ranking based on mean opinion scores (MOS) to measure perceptual quality. Our results obtained an average PSNR of $25.44$ dB in the full output images of the test set. This is, $0.18$ dB above the $2^{nd}$ place and $1.35$ dB below the best PSNR value in the Fidelity track.

Figures \ref{fig:comparison1} and \ref{fig:comparison2} show examples of our best results for the Fidelity and Perceptual tracks on images used for validation during our training process. These images show the values of RMSE (measuring fidelity) as well as the Perceptual index proposed in \cite{PIRM-SR} to objectively measure perceptual quality. Overall, the results are consistent with the perception/distortion trade--off in \cite{Blau_2018_CVPR}. For the images \texttt{142} and \texttt{622} we observe that our results for the Perceptual track achieve a Perceptual index better than original images. In image \texttt{622} we attribute this result to a blurred background in the original image, that our system shows more focused and with sharper features. In image \texttt{622} the results seem to reflect a failure of the Perceptual index as the original image clearly displays a face that our output shows with ambiguous characteristics. This shows that our system does not go as far as to fake people's identities, but it seems to correctly identify a person by adding skin and body textures. For images \texttt{472}, \texttt{808} and \texttt{1466}, the Perceptual indexes are clearly below those of the original image. According to our subjective evaluation, we observe clear differences in the fine level features like: tree leaves and water for image \texttt{472}; hair, eyebrows and eyelashes in \texttt{808}; and textbook spines in \texttt{1466}. From a far away look these details become less perceptible, indicating that the Perceptual index correlates better with a close distance observer. This is probably caused by the resolution of example images used to adjust the Perceptual index, that are much smaller than 8K.

Figure \ref{fig:interpolation} shows the effect of continuously adding noise in the input of our model for the Perceptual track. We observe that our model has correctly interpreted the target of our loss function, aiming at high fidelity with zero noise amplitude and high perceptual quality for unit noise amplitude. The trajectory in the perception distortion plot is mostly concave, except for the high perception corner. This trajectory is not meant to be optimal as perception/distortion optimality is far from being represented in the loss function. Our loss function only considers the corner cases of zero and unit noise amplitudes. The transition from high fidelity to high perceptual quality is smoother than similar tests performed with MGBP in \cite{G-MGBP}. The plot for the contextual similarity, as defined in \cite{mechrez2018contextual}, shows the best result in the beginning of the strong transition from high fidelity to high perceptual quality. Before this point it behaves as the perceptual index (improving as noise increases) and after this point it changes to a behavior closer to a fidelity score (worsening as noise increases). Compared to similar experiments performed in \cite{G-MGBP}, the contextual loss here seems to be more focused on preventing low fidelity scores.

Finally, Figure \ref{fig:visualization} shows interpretability results obtained by using the Deep Filter Visualization (DFV) method from \cite{PNavarrete_2019a}. To perform this complex analysis on a large model such as MGBPv2, with more than $20$ million parameters, we use the so--called \emph{Linearscope} method recently introduced in \cite{LinearScopes}. For a given pixel in the input image (blue circles on the left side), Figure \ref{fig:visualization} displays the impulse response for the network model with frozen activations (all ReLU's acting as if the input image did not change). This represents the equivalent to an upscaling filter that adapts to the pixel location. In flat areas (example at the bottom of Figure \ref{fig:visualization}) the upscaling filter looks isotropic and similar to a bicubic upscaler. In other locations, the filter strongly follows edges in hair and fingers, with receptive fields that extend for several hundred pixels. Overall, this confirms that the system has learned the geometry of the content.

\section{Conclusions}
The second generation of Multi--Grid Back--Projection (MGBPv2) networks improves its predecessor by: making better use of the multigrid recursion; simplifying its modules; and redesigning training and inference strategies to improve scalability. MGBPv2 achieves state--of--the--art performance by winning the Perceptual track of the ICCV AIM Extreme-SR Challenge 2019.

{\small
\bibliographystyle{ieee_fullname}
\bibliography{bibliography}
}

\end{document}